\documentclass[reprint,twocolumn]{revtex4}
\usepackage{amsfonts,amsmath,amssymb}
\usepackage{graphicx,color}

\newcommand{\MCS}{\textnormal{MCS}}

\begin{document}

\title[Membranes under Tension]
{Coarse-Grained Simulations of Membranes under Tension}

\author{J\"org Neder}
\email{joerg.neder@uni-konstanz.de}
\affiliation{Department of Physics, University of Konstanz, Germany}

\author{Beate West}
\affiliation{Department of Physics, University of Bielefeld, Germany}

\author{Peter Nielaba}
\affiliation{Department of Physics, University of Konstanz, Germany}

\author{Friederike Schmid}
\affiliation{Institute of Physics, University of Mainz, Germany}

\begin{abstract}

We investigate the properties of membranes under tension by
Monte-Carlo simulations of a generic coarse-grained model for lipid
bilayers. We give a comprising overview of the behavior of several
membrane characteristics, such as the area per lipid, the monolayer
overlap, the nematic order, and pressure profiles. Both the
low-temperature regime, where the membranes are in a gel $L_{\beta'}$
phase, and the high-temperature regime, where they are in the fluid
$L_{\alpha}$ phase, are considered. In the $L_{\beta'}$ state, the
membrane is hardly influenced by tension. In the fluid state, high
tensions lead to structural changes in the membrane, which result in
different compressibility regimes. The ripple state $P_{\beta'}$,
which is found at tension zero in the transition regime between
$L_{\alpha}$ and $L_{\beta'}$, disappears under tension and gives way
to an interdigitated phase.  We also study the membrane fluctuations
in the fluid phase.  In the low tension regime the data can be fitted
nicely to a suitably extended elastic theory. At higher tensions the
elastic fit consistently underestimates the strength of
long-wavelength fluctuations.  Finally, we investigate the influence
of tension on the effective interaction between simple transmembrane
inclusions and show that tension can be used to tune the hydrophobic
mismatch interaction between membrane proteins.
\end{abstract}

\maketitle

\section{Introduction} 

Biological membranes are made of lipid bilayers with incorporated
proteins. These barriers define the inside and the outside of a cell,
separate the functional compartments in cells, and are indispensable
for life \cite{bertym:02}.  The microscopic surface tension of
membranes is usually small or vanishes altogether \cite{nelpir:04},
but there may also be situations, where membranes are under
considerable stress due to osmotic pressure differences.  For example,
epithelial cells exposed to transmembrane osmotic gradients can be
expected to develop a state of tension under physiological conditions
\cite{sovmac:97}.  Similarly, osmotically induced tension may play a
decisive role during conformational changes, fission or fusion of
cells \cite{shilip:05,grashi:07}. Another situation where membranes
experience stress is under the influence of ultrasonic pulses
\cite{mit:05}.  Applied perpendicular to a lipid membrane, shock
pulses can promote significant structural changes similar to those
induced by lateral tension. The effect of such pulses on membranes is
of considerable medical interest.  In this context Koshiyama \emph{et
al.} have studied phospholipid bilayers under the action of a shock
wave in atomistic molecular dynamics simulations \cite{koskod:06}.

Despite the advances in computer technology throughout the past decades,
atomistic modeling of lipid bilayers on length scales of a few nanometer still
requires huge computing resources or even goes beyond the current capabilities
of high performance architectures. This motivates the use of coarse-grained
models.  They can give fundamental insights into the physics of a certain
system, and, moreover, they provide powerful tools for the interpretation of
the behavior of complex systems, like lipid membranes
\cite{vot:09:1,muekat:06,des:09,sch:09,venspe:06}. 

The aim of the work presented here is to study the effect of an externally
applied tension on the physical properties of a model bilayer, and on the
behavior of incorporated model proteins.  We employ a generic coarse-grained
model of amphiphiles developed in a top-down approach\cite{lensch:05}. For
tensionless systems this model has already been used very successfully to
reproduce various bilayer phases including the symmetric and asymmetric
"rippled" $P_{\beta'}$ states \cite{lensch:07,wessch:10} and to study 
membrane-protein interactions \cite{wesbro:09}.
Recent simulations on membranes under mechanical stress have often dealt with
the formation, structure and stability of hydrophilic pores
\cite{tieleo:03,leomar:04,coodes:05} or with the influence of surface tension
on transmembrane channel stability and function \cite{gulsch:04,zhuvau:05}. In
this paper, we will primarily be concerned with the structural changes of pure
membranes in response to lateral stresses, focussing on unporated systems.
Since our model exhibits a rather realistic phase behavior of the model
membrane at different temperatures, we can study different membrane states,
{\em i.e.}, the liquid, the ripple, and the gel state

Our paper is organized as follows: First we describe the underlying lipid model
and outline the simulation methods. Then the simulation results are presented
starting with a phenomenological introduction, where the effects of an external
tension on the model bilayer in different phases are described.  Thereafter a
quantitative analysis of these bilayers is performed and the characteristics of
the bilayers are examined with respect to their behavior under external
tension. In the last part we show the influence of tension on the effective
interaction potential of two simple model proteins. A brief summary concludes
our paper.
 
\section{Model and Methods} 

\label{sec:simulation}

In our model \cite{schdue:07} the lipid molecules are represented by
chains made of one head bead and six tail beads, and we have
additional solvent beads.  Hence we have three types of beads, $h,t$,
and $s$ for head, tail, and solvent beads. Within the lipid chains,
neighbor beads at a distance $r$ interact {\em via} a finite
extensible non-linear elastic (FENE) potential
\begin{equation} 
V_{\mathrm{FENE}}(r) = - \frac{1}{2} \epsilon_{\mathrm{FENE}} 
  (\Delta r_{\mathrm{max}})^2 
  \ln \left( 1 - \frac{(r - r_0)^2}{\Delta r_{\mathrm{max}}^2}
  \right), 
\end{equation}
with the spring constant $\epsilon_{\mathrm{FENE}}$, the equilibrium
bond length $r_0$, and the cutoff $\Delta r_{\mathrm{max}}$.
The angles $\theta$  between subsequent bonds in the lipid are
subject to a stiffness potential 
\begin{equation}
V_{\textnormal{BA}}(\theta) = \epsilon_{\textnormal{BA}} (1 - \cos \theta),
\end{equation}
with the stiffness parameter $\epsilon_{\textnormal{BA}}$.
Beads of type $i$ and $j$ which are not direct next neighbors in a chain 
interact {\em via} a truncated and shifted Lennard-Jones potential,
\begin{equation} 
V_{\textnormal{LJ}}(r/\sigma_{ij}) = \left\{ \begin{array}{cl} \epsilon
\left( \frac{\sigma_{ij}^{12}}{r^{12}} 
- 2 \frac{\sigma_{ij}^6}{r^6}\right) - V_{c,ij}, & \textrm{if}\  
r < r_{c,ij} \\ 0 & \textrm{otherwise.} \end{array}
\right. 
\label{equ:V_LJ}
\end{equation}
The offset $V_{c,ij}$ is chosen such that $V_{\textnormal{LJ}}(r/\sigma_{ij})$ 
is continuous at the cutoff $r_{c,ij}$. The parameter 
$\sigma_{ij} = (\sigma_i + \sigma_j)/2$ is the arithmetic mean of the 
diameters $\sigma_i$ of the interaction partners, and
$r_{c,ij} = 1\,\sigma_{ij}$ for all partners $(ij)$ except $(tt)$
and $(ss)$: $r_{c,tt} = 2\,\sigma_{tt}$ and $r_{c,ss} = 0$. Hence tail beads
attract one another, all other interactions are repulsive,
and solvent beads do not interact at all with each other.
This way of modeling the solvent environment (the so-called 'phantom solvent')
\cite{lensch:05} combines the advantages of explicit and implicit solvent
models: Like implicit solvent models, the solvent environment does not develop
any artificial internal structure, and it is very cheap (in Monte-Carlo or
Brownian dynamics simulations, less than 10 \% of the total computing time is
spent on the solvent beads). Like explicit solvent models, the model can be
used to study solvent-mediated phenomena such as the effect of hydrodynamic
interactions on membrane dynamics. This is much more difficult with
solvent-free models. We should note, however, that we mainly consider static
membrane properties in the present work, using Monte-Carlo simulations.

We use the model parameters \cite{schdue:07} $\sigma_h = 1.1 \,\sigma_t$, $r_0
= 0.7\,\sigma_t$, $\Delta r_{\mathrm{max}} = 0.2 \,\sigma_t$,
$\epsilon_{\mathrm{FENE}} = 100\,\epsilon/\sigma_t^2$, and
$\epsilon_{\textnormal{BA}} = 4.7\,\epsilon$. At the pressure $P =
2.0\,\epsilon/\sigma_t^3$, the model reproduces the main phases of
phospholipids, {\em i.e.}, a high-temperature fluid $L_\alpha$ phase at
temperature $k_B T > k_B T_m \sim 1.2\,\epsilon$ and a low-temperature tilted
gel ($L_{\beta'}$) with an intermediate modulated ripple ($P_{\beta'}$) phase
\cite{lensch:07}.  The energy and length scales can be mapped to SI-units
\cite{wesbro:09} by matching the bilayer thickness or, alternatively, the area
per lipid and the temperature of the main transition to those of DPPC, giving
$1\,\sigma_t \sim 6\,\textrm{\AA}$ and $1\,\epsilon \sim 0.36 \times
10^{-20}\,\mathrm{J}$.  The elastic properties of the membranes in the fluid
state were then also found to be comparable to those of DPPC membranes
\cite{wesbro:09}.
 
Unless stated otherwise, the systems were studied by Monte-Carlo simulations at
constant normal pressure $P_N = 2.0 \, \epsilon/\sigma_t^3$ and constant
temperature $T$ with periodic boundary conditions in a simulation box of
variable shape and size.  We put the tension into effect by an additional
energy term $-\Gamma A$ to the Hamiltonian of the system, where $A$ is the
projected area of the bilayer onto the $xy$ plane.  This alters the lateral
components of the pressure tensor within the membrane. The non-interacting
solvent particles, which probe the free volume and force the lipids to
self-assemble, are not affected by this additional energy contribution. They
ensure that the normal pressure $P_N$ is kept fixed at the required value.
Thus, we are performing Monte-Carlo simulations in an $N P_N \Gamma \, T$
ensemble with effective Hamiltonian
\begin{equation} 
\label{equ:hamiltonian_gamma}
H_{\mathrm{eff}} = E + P_N V - \Gamma A - N k_B T \ln(V/V_0)
\quad , 
\end{equation} 
where $E$ is the interaction energy, $V$ the volume of the simulation box,
$V_0$ an arbitrary reference volume and $N$ the total number of
beads. In contrast to the experimental situation, where the lateral pressure
of a lipid bilayer cannot be controlled very easily \cite{mou:05}, our
implementation of the tension is straightforward. Since we are in
full control of the lateral pressure, we can gain insight into states 
and structures of lipid bilayers by means of computer simulations, 
which are difficult to investigate in experimental setups.

In practice, two main types of Monte-Carlo moves were proposed and then
accepted or rejected according to a Metropolis criterion, namely 1)
translational local moves of the beads, and 2) global moves which change the
volume of the simulation box or its shape \cite{schdue:07}.  During
one Monte-Carlo step ($\MCS$) there is on average one attempt to move each
bead. Since the global moves require rescaling of all particle coordinates,
which is rather expensive from a computational point of view, they are
performed only every 50th $\MCS$ on average. 
The volume and shear moves are necessary to maintain the desired
surface tension. An advantage of our scheme is that the correct area per lipid
required by the external tension is adopted by construction. There is no need
for searching the required state by testing various values of fixed area per
lipid. 
We also implemented flip-flop moves with an inversion of the end-to-end 
vector of a complete lipid chain in its last tail bead, but found the 
acceptance rate of this 'molecular' move due to the
density of the bilayer far too small. 

In some cases it turned out to be more convenient to keep the box height $L_z$
fixed and vary only the planar extension given by $L_x$ and $L_y$. In this case,
the number of phantom solvent particles was allowed to fluctuate, {\em i.e.}, 
additional Monte-Carlo moves were attempted where solvent particles were 
removed from the system or randomly inserted (semi-grand canonical simulations).
The solvent chemical potential was set to
\begin{equation}
\mu_s = k_B \ln(V_0 P_N /k_B T)\,.
\end{equation}
Now, the semi-grand canonical Hamiltonian reads
\begin{equation} 
H_{\mathrm{s,gc}} =  E + P_N V - \Gamma A - \mu_s N_s -
k_B T\ln\left[\left(V/V_0)^N/N_s!\right)\right]\,,
\end{equation}
where $N_s$ is the momentary number of solvent particles in the system.

When the system sizes were very large (Sec.~\ref{sec:fluctuations}), the
simulations were carried out on parallel processors using a domain
decomposition scheme described in Ref.~\onlinecite{schdue:07}, which ensures
that every Monte-Carlo move fulfills detailed balance exactly (in most other
decomposition schemes proposed in the literature, detailed balance is violated
at the boundaries between the domains). We have checked by comparison with
single-processor simulations that the results were not affected by the
parallelization.

In this paper the results presented in Section~\ref{sec:pmf} were obtained
using the semi-grand canonical solvent model, and the diffusivity measurements
(Sec.~\ref{sec:diffusivity}) in the $NVT$-ensemble (in the latter case, only
local bead translations were attempted and all simulations were run on a single
processor). The other simulations used the $NP_N \Gamma \, T$-ensemble or.

The simulations of stressed membranes were carried out for a duration of at
least $4\times 10^6\,\MCS$, and we checked that the observed quantities did not
show any tendency for drift. In particular, we have examined in detail the
long-wavelength undulations of the membranes, which are presumably the slowest
relaxation modes in the system. After an equilibration time of $2\times
10^6\,\MCS$ the results from two successive runs of length $2\times 10^6\,\MCS$
were identical within the error. Hence the analyzed systems can be assumed to
be in a stable state.

\section{Simulation Results} 

\label{sec:results}

\begin{figure}
\begin{center}
\includegraphics[width=\columnwidth]
  {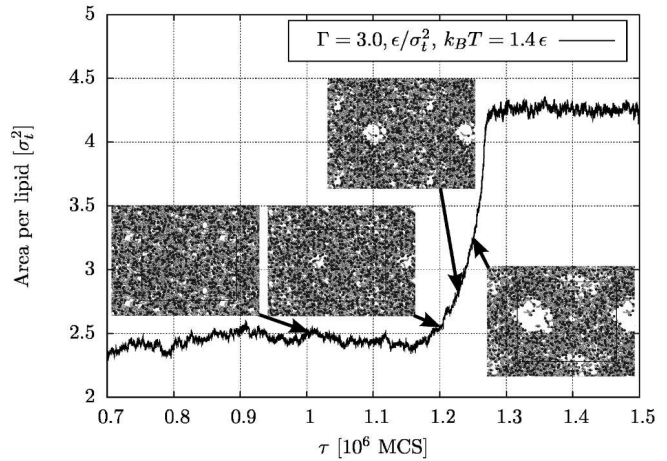}
\caption{Snapshots (top view) of a rupture event at $k_B T = 1.4\,\epsilon$ and
$\Gamma = 3.0\,\epsilon/\sigma_t^2$. The underlying curve shows the
corresponding evolution of the area per lipid. At about $4.3\,\sigma_t^2$, the
area per lipid levels off, because a restriction was imposed on the minimal
size of the simulation box in the $z$ direction.}
\label{fig:T1p4_20x20_g3p0-rupture_and_snapshots}
\end{center} 
\end{figure}

We have applied tensions of up to $\Gamma = 3.0\,\epsilon/\sigma_t^2$,
corresponding to $30\,\textrm{mN}/\textrm{m}$, which is close to the threshold
for rupture in the fluid phase in our model: The membranes remained stable up
to temperatures $k_B T = 1.3\,\epsilon$, but ruptured at $k_B T =
1.4\,\epsilon$ (see Fig.~\ref{fig:T1p4_20x20_g3p0-rupture_and_snapshots}).
This is consistent with atomistic simulations of Leontiadou \emph{et al.}, who
found a critical tension for stable porated bilayers of $\approx
38\,\textrm{mN}/\textrm{m}$ in the context of pore formation with an atomistic
model of DPPC \cite{leomar:04}.  In contrast, the bilayer in the gel phase can
sustain much higher tensions and stays stable for values of $\Gamma$ up to
$4.0\,\epsilon/\sigma_t^2$ and higher.  The areal expansion and reduction in
thickness of the bilayer is substantial. In our simulations, we observe an
increase in area per lipid of more than $40\,\%$  and a decrease in bilayer
thickness of more than $30\,\%$ for high tensions, where the systems can still
be found in non-rupturing configurations. 
In other simulation studies a comparable or even larger increase in the area
per lipid was observed under tension without rupturing of the bilayer. Groot
and Rabone report an areal expansibility of more than 70\% for mixed membranes
before these are finally ripped apart \cite{grorab:01}. Grafm\"uller \emph{et
al.} see an areal gain of 60\% under tension for one of their parameter sets
\cite{grashi:09}.  

\begin{figure}
\begin{center}
\includegraphics[width=\columnwidth]{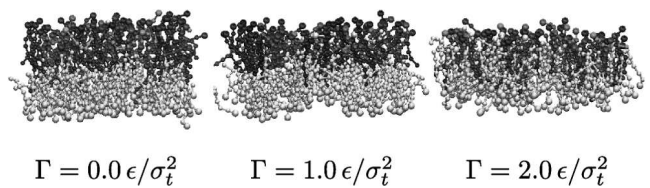}
\caption{Snapshots of bilayer configurations in the fluid phase at $k_B T =
1.3\,\epsilon$. Gray scale coding: Light gray molecules point upwards from
head to tail, dark gray molecules point downwards.
At increasing tension the two monolayers partly interdigitate.
The size of the beads are not to scale.}
\label{fig:snapshots_T1p3_10x10}
\end{center} 
\end{figure}

\begin{figure}
\begin{center}
\includegraphics[width=\columnwidth]{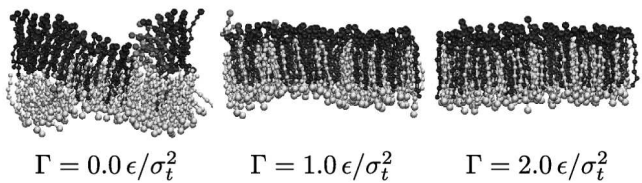}
\caption{Snapshots of bilayer configurations in the ripple phase at 
$k_B T = 1.2\,\epsilon$ (gray scale coding as in Fig.
\protect\ref{fig:snapshots_T1p3_10x10}).
On the left the tensionless system with a pronounced ripple is
depicted. This ripple vanishes as tension is applied, giving way
to a completely interdigitated structure (middle and right figure).
}
\label{fig:snapshots_T1p2_10x10}
\end{center}
\end{figure}

\begin{figure}
\begin{center}
\includegraphics[width=\columnwidth]{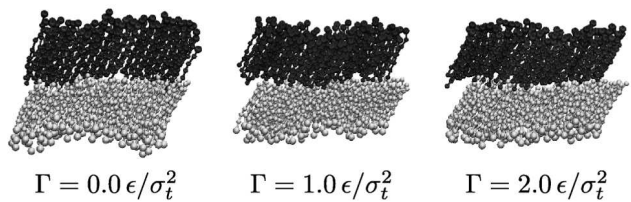}
\caption{Snapshots of bilayer configurations in the gel phase 
$k_B T = 1.1\,\epsilon$ (gray scale coding as in 
Fig.~\ref{fig:snapshots_T1p3_10x10}).
No interdigitation at increasing tension is observed.}
\label{fig:snapshots_T1p1_10x10}
\end{center} 
\end{figure}

We begin with giving a qualitative overview over the behavior observed in 
the different bilayer phases. Figs.~\ref{fig:snapshots_T1p3_10x10},
\ref{fig:snapshots_T1p2_10x10}, and \ref{fig:snapshots_T1p1_10x10}
show configuration snapshots of bilayers under tension 
at the temperatures $k_B T = 1.3\,\epsilon$, corresponding to a fluid state,
$k_B T = 1.2\,\epsilon$, corresponding to a rippled state, and
$k_B T = 1.1\,\epsilon$, corresponding to a tilted gel state.

In the fluid phase (Fig.~\ref{fig:snapshots_T1p3_10x10}), the membrane
stretches under tension, and the two monolayer leaflets become less well
separated (right). Thus high tensions change the structure of the membrane. 
This effect is even more pronounced in the rippled state 
(Fig.~\ref{fig:snapshots_T1p2_10x10}; here the original rippled structure 
was obtained by cooling tensionless equilibrated configurations from 
the fluid phase at $k_B T = 1.3\,\epsilon$ down to $k_B T = 1.2\,\epsilon$).  
Under tension, the ripple unravels and gives way to an interdigitated phase.  
In contrast, the gel state (Fig.~\ref{fig:snapshots_T1p1_10x10}) is hardly 
affected under tension. The two monolayers remain well-separated. Only the 
average lipid tilt away from the bilayer normal is slightly enhanced from 
$\langle \theta \rangle = 23^{\circ}$ in the tensionless state to 
$\langle \theta \rangle = 30^{\circ}$ at the tension 
$\Gamma = 3.0\,\epsilon/\sigma_t^2$ (snapshot not shown).

After this qualitative overview, we turn to a more quantitative 
analysis of the behavior of bilayers under tension.

\subsection{Global characteristics of pure membranes}

We first consider the area per lipid, which is obtained by dividing the
projected area of the bilayer in the $xy$ plane by half the number of lipids in
the system. We note that the difference between the number of lipids in the
upper and the lower monolayer was always very small. "Flip-flop" moves are
practically never observed for systems in the gel phase. In the fluid phase,
about $10\,\%$ of the lipids were exchanged between the monolayers during
$10^6\,\MCS$, but fluctuations of the average number of lipids in each
monolayer are still less than $2\,\%$.

We do not observe any dependence of the area per lipid on the system size. For
the temperature $k_B T=1.3\,\epsilon$, we have compared data from four
different system sizes, ranging from $200$ lipids to $7,200$ lipids; the
results were identical within the statistical error (\mbox{$\approx 1\,\%$}).
These results are in agreement with the findings of Marrink and Mark in
atomistic simulations \cite{marmar:01}, or with those of Kranenburg \emph{et
al.} \cite{kraven:03:2}, who studied a coarse-grained model of amphiphilic
surfactants by  a combined DPD and Monte-Carlo scheme, imposing the surface
tension in a way similar to ours.

Assuming that the area per lipid $A$ depends linearly on the applied tension
$\Gamma$, one can calculate the mean area compressibility modulus $k_A$ using
the relation $ k_A = A_0 {\Delta \Gamma}/{\Delta A}, $ where $A_0$ is the
equilibrium area of the tensionless membrane.  This yields the values
of $k_A$ listed in Table~\ref{tab:compressibility}.  Here only data from
configurations which remained stable for long simulation runs (up to
$10^7\,\MCS$) have been taken into consideration, {\em i.e.}, the data for the
state point $k_B T = 1.4\,\epsilon$, $\Gamma = 3.0\,\epsilon/\sigma_t^2$, which
lies beyond the rupture threshold, were omitted.  

\begin{table}
\begin{center}
\begin{tabular}{l||c|c|c|c}
$k_B T$ [$\epsilon$] 		& 1.0	& 1.1	& 1.3 	& 1.4 \\
\hline
$k_A$ [$\epsilon/\sigma_t^2$]	& 40	& 32	& 4.3	& 4.6 \\
\end{tabular}
\end{center}
\caption{Mean area compressibility modulus $k_A$ over the whole
range of tensions.
}
\label{tab:compressibility}
\end{table}

For membranes in the gel phase ($k_B T = 1.0\,\epsilon$ and $k_B T =
1.1\,\epsilon$), the number $k_A$ fully characterizes the behavior of the area
per lipid over the whole investigated range of tensions (data not shown). The
extensibility of the bilayer in the gel phase is significantly smaller than in
the fluid phase, and constant over the whole range of tensions under
investigation.  The most noticeable effect of the tension is the increase in
tilt angle of the lipids mentioned earlier, which leads to a slightly reduced
membrane thickness. 

\begin{figure}
\begin{center}
\includegraphics[width=\columnwidth]{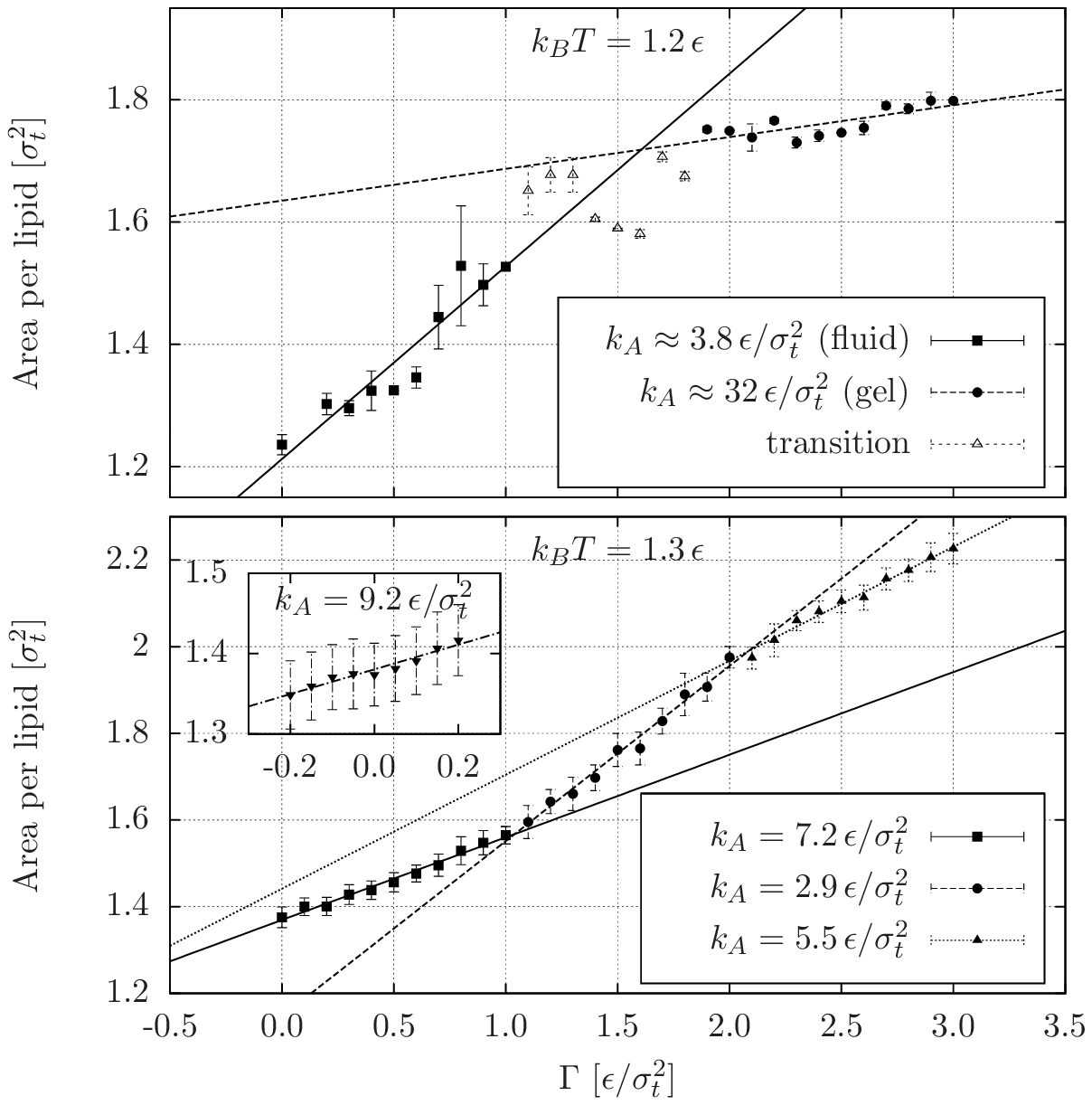}
\caption{Area per lipid $A$ vs. tension $\Gamma$ for two different temperatures
as indicated, together with linear fits to different compressibility regimes. \\
{\bf Top:} As the tension on a patch of bilayer in the ripple-phase
increases, the areal compressibility changes from fluid-like to gel-like.\\
{\bf Bottom:} In the fluid phase, three different compressibility regimes
are observed, $\Gamma = 0$--$1\,\epsilon/\sigma_t^2$,
$\Gamma = 1$--$2\,\epsilon/\sigma_t^2$, and 
$\Gamma = 2$--$3\,\epsilon/\sigma_t^2$.
The inset focuses on the tensionless membrane (see text for explanation)}
\label{fig:T1p2-1p3-area_per_lipid-fit_ranges}
\end{center}
\end{figure}

The behavior of membranes in the fluid or ripple state is more 
complicated. Fig.~\ref{fig:T1p2-1p3-area_per_lipid-fit_ranges}
shows the corresponding data for the area per lipid as a function
of the applied tension.

We first discuss the situation at the temperature $k_B T = 1.2\,\epsilon$,
where the tensionless membrane is in the ripple phase. As we have already
discussed earlier (Fig.~\ref{fig:snapshots_T1p2_10x10}), the membrane undergoes
a phase transition to an interdigitated phase under tension.
Fig.~\ref{fig:T1p2-1p3-area_per_lipid-fit_ranges}, top, shows that this is
associated with a pronounced change of the compressibility.  Up to values of
about $1.0\,\epsilon/\sigma_t^2$, the area per lipid increases steeply, the
resulting value of $k_A$ is even lower than that of the fluid phase.  At high
tensions, $\Gamma = 1.9\,\epsilon/\sigma_t^2$ and above, the approximate areal
compressibility is strongly enhanced and comparable to values obtained for
systems in the gel phase.  Thus the extensibility of the bilayer changes from
fluid-like to gel-like under tension.  

In the fluid phase (Fig.~\ref{fig:T1p2-1p3-area_per_lipid-fit_ranges}, bottom),
the tension-induced structural changes in the membranes are less dramatic, but
they can still be associated with compressibility changes. The data shown in
Fig.~\ref{fig:T1p2-1p3-area_per_lipid-fit_ranges}, bottom, suggest a
subdivision into three different compressibility regimes: Under tension, the
membrane switches from a less compressible low-tension state to a more
compressible high-tension state {\em via} a highly compressible intermediate.
Thus the structural changes in the membrane, which were observed in
Fig.~\ref{fig:snapshots_T1p3_10x10}, seem to be related to a crossover between
different membrane states and possibly even reflect the vicinity of a hidden
phase transition. 

The inset in Fig.~\ref{fig:T1p2-1p3-area_per_lipid-fit_ranges}, bottom, focuses
on the limit of very small tension/compression. In recent work, $k_A$ has been
determined for the case $\Gamma = 0.0\,\epsilon/\sigma_t^2$ with an alternative
method, {\em i.e.}, the detailed analysis of the fluctuation spectrum of
tensionless membranes \cite{wesbro:09}. Here, we have extracted the areal
compressibility modulus of the fluid membrane close to tension zero by both
compressing and extending the system slightly within the range of $\Gamma =
-0.2$ to $0.2 \,\epsilon/\sigma_t^2$. The resulting value for $k_A$ divided by
the square of the mean tensionless monolayer thickness $t_0\approx 3\,\sigma_t$
agrees with the value $k_A/t_0^2 = 1.1 \pm 0.2\,\epsilon / \sigma_t^2$ obtained
independently from the fluctuation analysis (see
Table~\ref{tab:elastic_parameters}).

\begin{figure}
\begin{center}
\includegraphics[angle=0,width=\columnwidth]{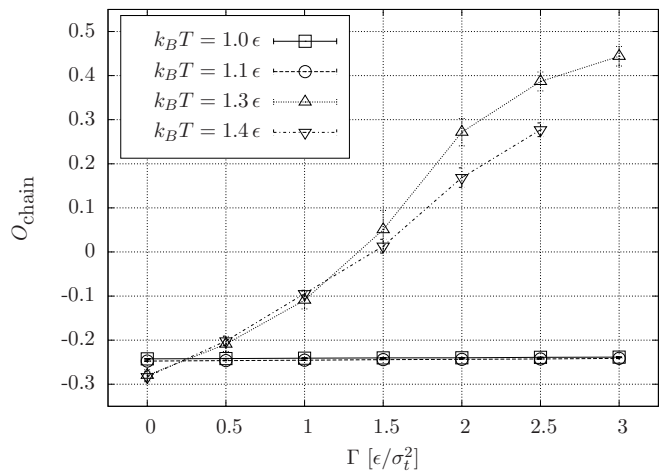}
\caption{Average overlap parameter (see text for definition) vs. tension.
Negative values indicate spatially well-separated monolayers. 
For interdigitated bilayers the overlap parameter is positive.}
\label{fig:T1p0-1p4_10x10-overlap}
\end{center}
\end{figure}

To characterize the global structure of the membrane in more detail,
we next consider the interdigitation of the two monolayers. It can
be characterized in terms of the 'overlap parameter'
$O_{\textrm{chain}} = \langle 2 (l_z - t_0)/l_z \rangle$, 
originally introduced by Kranenburg \emph{et al.} \cite{kraven:03:2}.
The results are shown in Fig.~\ref{fig:T1p0-1p4_10x10-overlap}.
They quantify the behavior which was already apparent from the snapshots shown
earlier (Figs.~\ref{fig:snapshots_T1p3_10x10}--\ref{fig:snapshots_T1p1_10x10}).  
In the gel phase, where we did not notice any overlap of the
monolayers even under tension $O_{\textrm{chain}}$ is always negative.
In the fluid phase, it becomes positive at high tensions.

\begin{figure}
\begin{center}
\includegraphics[width=\columnwidth]{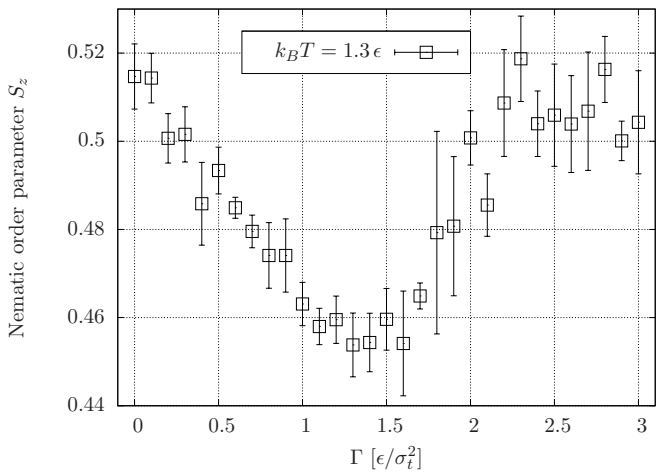}
\caption{Nematic order parameter of chains $S_z$ vs. tension.}
\label{fig:T1p3_20x20_g0-3-S_z}
\end{center} 
\end{figure}

Interestingly, this leads to a non-monotonic behavior of the orientational
chain order in the fluid phase, {\em i.e.}, the nematic order parameter $S_z =
1/2\langle (l_z/l)^2 - 1\rangle$.  Here, $l_z$ is again the $z$ component of
the end-to-end vector of a lipid chain, and $l$ its full length. As shown in
Fig.~\ref{fig:T1p3_20x20_g0-3-S_z}, the nematic order first decreases in the
low-tension regime $\Gamma < 1.0\,\epsilon / \sigma_t^2$.  At $ \Gamma \sim 1.5
- 2.0\,\epsilon / \sigma_t^2$, a rather steep increase ensues, followed by a
plateau in the high-tension regime $\Gamma > 2.0\,\epsilon / \sigma_t^2$.
Under tension, the lipids thus first disorder and tilt away from the bilayer
normal, which leads to an unfavorable packing in the hydrophobic bulk of the
membrane.  As interdigitation sets in, the lipids relax and assume once again
their preferred order.

\subsection{Pressure Profiles}

After having discussed these global properties of the membranes,
we now investigate the effect of the applied {\em external} stress
on the {\em internal} stress distribution inside the membrane. 
Stress distribution profiles influence, {\em e.g.}, the permeability
of membranes with respect to small molecules. To study them, we
have recorded the interfacial tension (or negative stress) profiles
\begin{equation}
\gamma(z) = P_{zz}(z) - \frac{1}{2}(P_{xx}(z) + P_{yy}(z))
\end{equation} 
in small systems of 200 lipids. The pressure tensor $P_{\alpha\beta}$ 
is obtained using the virial theorem,
\begin{equation}
P_{\alpha \beta} = \frac{N k_B T}{V} \delta_{\alpha\beta} 
  + \frac{1}{V} \left \langle  \sum_i r_i^{\alpha} F_i^{\beta}
\right \rangle .
\end{equation}
Here, $\mathbf{r}_i$ is the position of particle $i$, $\mathbf{F}_i$
the force acting on this particle, $N$ the number of particles, $T$
the temperature, and $V$ the volume. 
The local distribution of the pressure along the bilayer normal was
obtained by dividing the system into 50 vertical slabs and distributing
the pressure contributions onto these slabs according to the convention
of Irving and Kirkwood \cite{irvkir:50}. 
The pressure profiles can also be used to cross-check the consistency of
our approach in Eq.~\ref{equ:hamiltonian_gamma}, since the
integral 
\begin{equation}
\Gamma_0 = \int \mathrm{d}z \: \gamma(z)
\end{equation}
has to match the externally applied tension. We have checked that this
was the case in all simulations.

\begin{figure}
\begin{center}
\includegraphics[angle=0,width=\columnwidth]{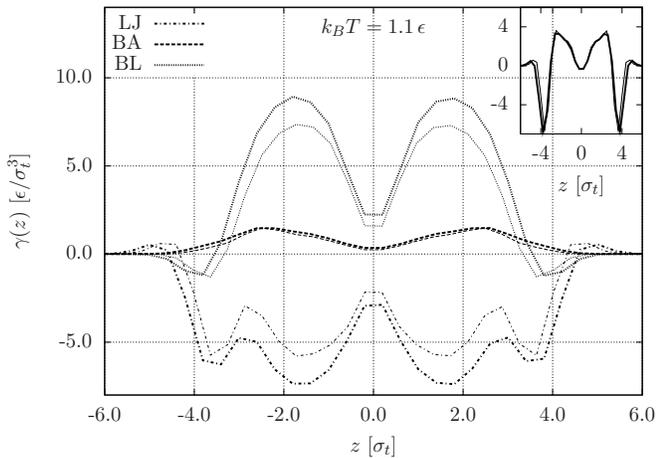}
\caption{Contributions of the Lennard-Jones (LJ),
bond-angle (BA), and bond-length (BL) interactions to the total stress
profiles in the gel phase, for the tensionless case (thick lines) and 
for the tension $\Gamma = 2.0\,\epsilon/\sigma_t^2$ (thin lines). 
The inset shows the total stress profile, {\em i.e.}, the sum of all
contributions, for $\Gamma = 0.0\,\epsilon/\sigma_t^2$ (thick line) and 
$\Gamma = 2.0\,\epsilon/\sigma_t^2$ (thin line).}
\label{fig:T1p1_10x10_g0-2-contr}
\end{center}
\end{figure}

\begin{figure}
\begin{center}
\includegraphics[angle=0,width=\columnwidth]{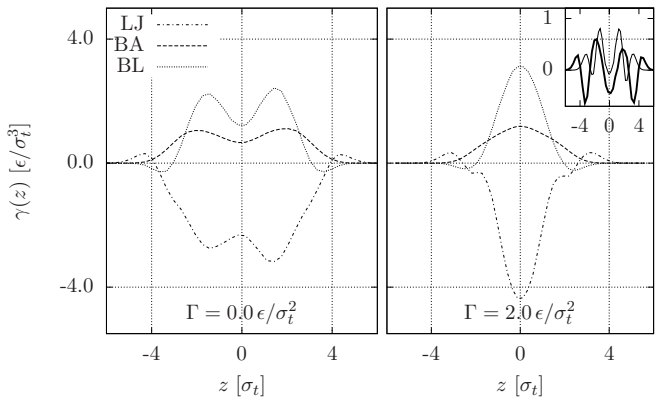}
\caption{Contributions of the Lennard-Jones (LJ),
bond-angle (BA), and bond-length (BL) interactions to the total stress
profiles in the fluid phase, in the tensionless case (left), and for the
tension $\Gamma = 2.0\,\epsilon/\sigma_t^2$ (right). The inset shows
the total stress profile for the tensionless bilayer (thick
line) and for $\Gamma = 2.0\,\epsilon/\sigma_t^2$ (thin line).}
\label{fig:T1p3_10x10_g0-2-contr}
\end{center}
\end{figure}

Pressure or stress profiles have been reported in various other studies
\cite{goelip:98,bralin:06,gulsch:04} for both mesoscopic and atomistic
simulations. We will briefly summarize the characteristic features of the total
stress profiles and analyze their change under tension in our model. The
behavior of the pressure profiles can be attributed to different interactions: 

The first positive peak (insets in Figures~\ref{fig:T1p1_10x10_g0-2-contr} and
\ref{fig:T1p3_10x10_g0-2-contr}) arises due to the purely repulsive
interactions of the head and the solvent beads.  Here a zone depleted from
(solvent) beads is formed and the head beads are effectively squeezed together
in the lateral direction.  The first negative peak in the head region indicates
that the head-head interactions are purely repulsive, pushing the system
laterally apart in this section of the bilayer.  The second positive peak,
located in the tail region, results from the interplay of the attractive
Lennard-Jones interaction between the tail beads and the contribution of
bond-angle and bond-length potentials. Interestingly, the contribution of the
(attractive) Lennard-Jones (LJ) interactions between tail beads is negative:
The net attraction between tail beads is stronger in the normal direction than
in the lateral direction. This is compensated and outbalanced by the
contributions of the bending of the lipid segments and the stretching of bonds,
which lead to positive tension in total.  Thus the stress in the hydrophobic
portion of the membrane is mostly sustained by intrachain interactions and not,
as one might expect, by the attractive Lennard-Jones interactions.  The
negative peak in the midplane of the bilayer originates from the absence of
intrachain interactions in this region, thus the effect of the Lennard-Jones
interactions takes over and the monolayers are effectively glued to each other.

Under the influence of an external tension, we observe a narrowing and shift of
the whole profiles, in qualitative agreement with previous atomistic
simulations by Gullingsrud and Schulten \cite{gulsch:04}.
In the gel phase, which is already exposed to very high internal stress, the
relative effect of the external tension is small. The inspection of the
different contributions to the local pressure shows that the external tension
leads to a decrease both of the Lennard-Jones (LJ) contribution and the bond
length (BL) contribution. The reduction of the LJ-contribution is higher,
leading to the observed shift. Due to the stiffness of the lipids in the gel
phase, the bond-angle contribution to the pressure profiles remains practically
unaltered. The change in the overall structure of the pressure profiles is also
only small. The slight change in membrane thickness results in a shift of the
outer peaks of the profiles towards the midplane ($z = 0$).

The effect of external tension is considerably more dramatic in the fluid phase
at $k_B T = 1.3\,\epsilon$. Although the absolute peak values of the pressure
profiles are reduced by a factor of about 8 to 10, the relative shifts are
much more pronounced.  First, the shift of the outermost peaks towards the
midplane is much higher, due to the fact that the membrane thickness decreases
more strongly under tension in this phase. Second, the shapes of the individual
contributions to the total profile change qualitatively under stress,
reflecting the structural change from a well-separated bilayer to an
interdigitated structure.  At high $\Gamma$, the individual terms have a simple
structure with single, positive or negative, peaks at the center of the
membrane. Nevertheless, the total stress profile still exhibits the oscillatory
features described above.

\subsection{Correlation functions and diffusion of lipids}

\label{sec:diffusivity}

\begin{figure}
\begin{center}
\includegraphics[angle=0,width=\columnwidth]{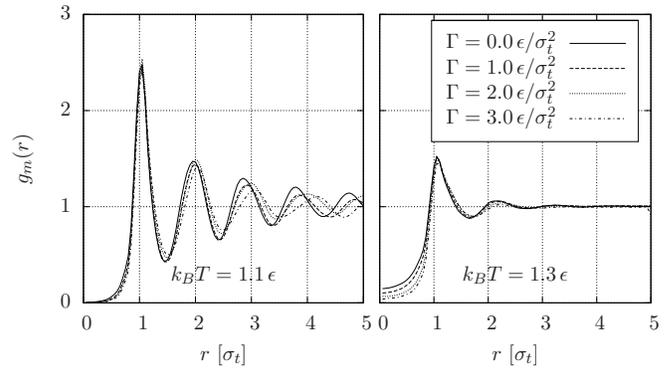}
\caption{Two-dimensional radial distribution function $g_m(r)$ of the
head groups within one monolayer
at $k_B T = 1.1\,\epsilon$ (left) and at 
$k_B T = 1.3\,\epsilon$ (right).}
\label{fig:T1p1-1p3_20x20_g0-3-rdf2d_hh_g_m}
\end{center}
\end{figure}

Next we consider the effect of the tension on the lateral structure 
of the bilayers and on the dynamic mobility of individual lipids.
Fig.~\ref{fig:T1p1-1p3_20x20_g0-3-rdf2d_hh_g_m} shows the two-dimensional
radial distribution function  $g_m(r)$ of the $(x,y)$-coordinates of
the head groups in the gel and the fluid state for different applied
tensions $\Gamma$. In the gel state, $g_m(r)$ exhibits a series of 
pronounced peaks, reflecting a high degree of order 
(Fig.~\ref{fig:T1p1-1p3_20x20_g0-3-rdf2d_hh_g_m}, left). Under tension, 
the higher order peaks shift to slightly larger distances, reflecting the 
enhancement of the area per lipid. The fluid membrane is much less structured. 
The head-head radial correlation functions within one monolayer decay
rapidly already at zero tension, and all higher order peaks disappear at
higher tension (Fig.~\ref{fig:T1p1-1p3_20x20_g0-3-rdf2d_hh_g_m}, right).  
In sum, the influence of the tension on the lateral structure of the
membranes is found to be largely negligible.

The situation is different for the diffusion coefficient of lipids. 
Although Monte-Carlo simulations do not provide an intrinsic timescale, one can
still obtain valuable information on the diffusional behavior of the lipids
from $NVT$ Monte-Carlo simulations that employ only local bead moves. In our
diffusivity studies, the initial configurations were taken from systems 
equilibrated in the $NP_N\Gamma\,T$-ensemble, but we measured the diffusivity 
of lipids in the $NVT$-ensemble, {\em i.e.}, no volume or shear moves were 
carried out during the simulation. We have monitored the pressure tensor 
for the duration of the diffusion measurements to check that it stayed
constant during the simulation. 

In the following, the basic "time unit" is one Monte-Carlo step
($\MCS$, the Monte-Carlo 'time scale'), and the "time" $\tau$ counts the number of
$\MCS$ since the start of the simulation.  We consider the lateral diffusion constant
within the membrane (or, more specifically, the projection of the membrane into
the ($xy$ plane)), defined as~\cite{goelip:98} 
\begin{equation}
\label{eq:lateral_diffusion_coefficient}
D_{xy}(\tau) = 
\frac{ \sum_i \left( \sum_{t=0}^\tau \Delta\mathbf{r}_i^{(xy)}(t) \right)^2}
{6 N \tau}
\,.
\end{equation} 
Here the sum $i$ runs over all lipid heads, $\mathbf{r}_i^{(xy)}$ denotes the
position of a lipid head $i$ in the $xy$ plane with respect to the center of
mass of the bilayer, and $\Delta \mathbf{r}_i^{(xy)}$ its difference from one
$\MCS$ time step to the next {\em without} the offset that has to be added if the 
head crosses the periodic boundaries. 

\begin{figure}
\begin{center}
\includegraphics[angle=0,width=\columnwidth]{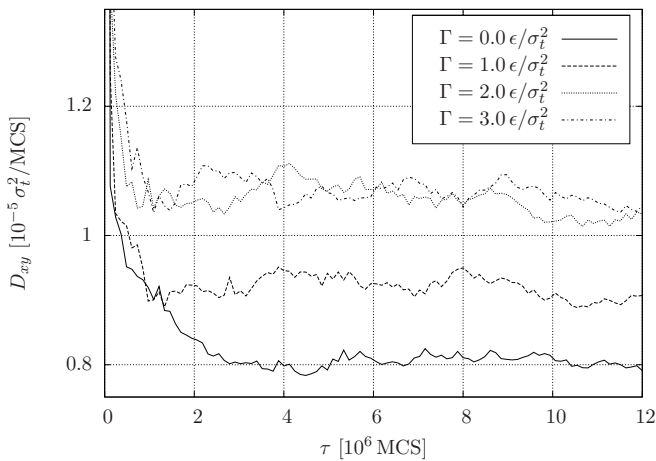}
\caption{Lateral diffusion coefficient $D_{xy}$ for different
tensions $\Gamma$. }
\label{fig:diffusion_xy}
\end{center}
\end{figure}

In the gel phase (data not shown), no diffusion was observed
over the whole length of the simulation 
($\tau_{\textnormal{max}} = 1.2 \times 10^7\,\MCS$).
The lipids basically fluctuate around their average positions.
The width of the fluctuations, 
$\sqrt{\langle {r^{(xy)}}^2 \rangle - \langle \mathbf{r}^{(xy)} \rangle^2} \sim 0.04\,\sigma_t$,
does not depend on the applied tension. The data for the fluid
phase are shown in Fig.~\ref{fig:diffusion_xy}. Here the lipids
diffuse freely, and the lateral diffusion constant $D_{xy}$ increases 
significantly if one applies moderate tensions up to 
$\Gamma = 2.0\,\epsilon/\sigma_t^2$ (Fig.~\ref{fig:diffusion_xy}). 
Interestingly, it does not increase further in the high-tension
regime, beyond $\Gamma = 2.0\,\epsilon/\sigma_t^2$, even though the 
area per lipid and the chain overlap parameter $O_{\textrm{chain}}$ 
are not yet saturated. 

It should be noted that the in-plane diffusion constant discussed here differs
from the true diffusion constant in the membrane, due to the presence of
membrane undulations (see also Sec.~\ref{sec:fluctuations}). The thermally
induced buckling of the membrane can lead to a substantial out-of-plane
component to the lipids' diffusional motion, which is not captured in our
definition of $D_{xy}$ in Eq.~\ref{eq:lateral_diffusion_coefficient}.  Various
theoretical studies have addressed this problem
\cite{gushal:97,najbro:07,reilei:07}. They commonly conclude that membrane
fluctuations lower the measured in-plane diffusion coefficient. As the tension
increases, the fluctuations of the membrane are suppressed (see
Sec.~\ref{sec:fluctuations}). Therefore, one may speculate that the increase of
the apparent diffusivity under tension measured in our simulations is, at least
in part, caused by the reduction of thermal bilayer undulations. However, this
is not sufficient to fully explain the 20\%-effect observed in
Fig.~\ref{fig:diffusion_xy}.

\subsection{Fluctuation spectra}

\label{sec:fluctuations}

To conclude the analysis of pure membranes, we study the thermal membrane
fluctuations, which were already mentioned in the previous section.
Theoretical considerations \cite{hel:78} and experiments \cite{isrwen:92} have
shown that lateral tension on fluid bilayers leads to suppression of thermal
fluctuations, which in turn decreases steric repulsion of vesicles and changes
adhesive properties. Moreover, the Fourier spectra of the height and thickness
fluctuations provide information on the elastic properties of the membranes.
Therefore, it is instructive to look at the development of undulation,
peristaltic and protrusion properties of our model membrane under tension.

To analyze our data, we use an extension of an elastic theory by Brannigan and
Brown \cite{brabro:06}, which we have already applied with success to the case
of the tensionless membrane \cite{wesbro:09}.  We shall not rederive the theory
here, but merely sketch the main assumptions. Brannigan and Brown describe a
planar fluctuating membrane as a system of two coupled monolayers surfaces
$h_{1,2}(x,y)$, of with each is characterized by two independent fields,
$z_{1,2}(x,y)$ and $\lambda_{1,2}(x,y)$ ($h = z + \lambda$), one accounting for
slow 'bending modes' and one for fast 'protrusion modes'\cite{goegom:99}.  They
furthermore make a number of approximations, which amount to the assumptions
that 
\begin{itemize}
\item[(i)] the protrusion modes of the two monolayers are independent 
degrees of freedom,
\item[(ii)] the bending modes $z_{1,2}(x,y)$ can be rewritten in terms of their
sum and their difference ($z_{\pm} = (z_1\pm z_2)/2$), corresponding to
(bending) height and (bending) thickness modes of the membrane, which in turn
decouple.  The (bending) thickness modes are subject to the constraint that the
volume per lipid is conserved.
\end{itemize}
With these assumptions, Brannigan and Brown construct a free energy functional
which is quadratic in the fluctuations of the fields $z$ and $\lambda$ (higher
order contributions are neglected) and can be used to calculate the thermally
averaged fluctuations of the total membrane height, $h=(h_1+h_2)/2$, and
monolayer thickness, $t=(h_1-h_2)/2$.

In our simulations, the situation is different from that considered by
Brannigan and Brown in two respects. First, we apply an external tension.
Second, we do not have well-separated monolayers, especially at high tensions.
We argue that the second point is not critical: If we associate the fields
$h_{1,2}$ with the positions of the {\em head group layers} layers rather than
whole {\em monolayers}, the assumptions enumerated above are still reasonable
and we obtain the same theory.  The first point is more subtle, since the
external tension is not an intrinsic material parameter such as, {\em e.g.},
the bending energy (which drives the bending fluctuations). 

To assess the effect of an applied external tension on the height fluctuations,
we first consider a simplified case, where the membrane is characterized by a
single surface manifold $h(x,y)$ (shifted to $\langle h \rangle = 0$) with
fixed surface area $A_s$ and variable projected area $A$. The surface area is
related to the projected area {\em via}
\begin{equation}
A_s = \int_{A} \mathrm{d}^2 r \sqrt{1+(\nabla h)^2} 
\approx A + \frac{1}{2} \int \mathrm{d}^2 r (\nabla h)^2 \,.
\end{equation}
Upon applying an external tension $\Gamma$, the free enthalpy $G$
of the system is given by 
\begin{equation}
\label{eq:gamma_1}
G = - \Gamma A 
\approx - \Gamma A_s + \frac{\Gamma}{2} \int \mathrm{d}^2 r \: (\nabla h)^2,
\end{equation}
where \mbox{$\Gamma A_s$} is a constant. Hence the external tension couples to
the fluctuations of the local membrane height $h(x,y)$ in the same way as an
internal interfacial tension couples to the fluctuations, {\em e.g.}, a
gas-liquid interface \cite{rowwid:82}.

The same type of argument can be applied to the model of Brannigan and Brown,
where the membrane has finite thickness and the lipid volume, rather than the
lipid area, is conserved: Let $z_+(x,y)$ and $2z_-(x,y)$ denote the local
membrane height and thickness as before, with $\langle z_+ \rangle = 0$. Let
furthermore $D(x,y)$ denote the true local membrane thickness, evaluated with
respect to the local surface normal, {\em i.e.}, $z_- (x,y)= D \sqrt{1+(\nabla
z_+)^2}/2$.  The thickness is taken to fluctuate weakly about its mean value
$\overline{D}$, {\em i.e.}, $D(x,y) = \overline{D}+\delta(x,y)$.  The number of
lipids on the projected area element $\mathrm{d}A$ is then given by 
\begin{eqnarray}
N &=& \frac{2}{v} \int_{A} \textnormal{d}^2 r \: z_- 
= \frac{1}{v} \int_{A}\mathrm{d}^2 r \: \sqrt{1 + (\nabla z_+)^2} \: D 
\nonumber \\
&\approx&
\frac{1}{v} \left(
 \overline{D} A + \frac{\overline{D}}{2} \int \mathrm{d}^2 r (\nabla h)^2
+ \int \mathrm{d}^2 r \: \delta
\right)
\end{eqnarray}
with the lipid volume $v$, where we have expanded up to second order in the
fluctuating fields $h$ and $\delta$. Upon applying an external tension
$\Gamma$, the free enthalpy acquires an additional term
\begin{equation}
\label{eq:gamma_2}
G_{\Gamma} = 
- \Gamma A =
-\frac{\Gamma v N}{\overline{D}} 
+ \frac{\Gamma}{2} \int_{A} \mathrm{d}^2 r (\nabla z_+)^2
+ \frac{\Gamma}{\overline{D}} \int \mathrm{d}^2r \: \delta,
\end{equation}
where $\Gamma v N/\overline{D} = \mbox{const.}$ Hence the external tension
again has the same effect on the height fluctuations than an internal tension.
The last term in Eq.~(\ref{eq:gamma_2}) results in an effective thinning of the
membrane and does not contribute to the fluctuation spectra in the quadratic
order considered here.

Supplementing the free energy of Brannigan and Brown~\cite{brabro:06}
with this additional tension term, we obtain the Hamiltonian
(in Fourier space)
\begin{equation}
\begin{split}
F =& \frac{1}{2} \sum_{\mathbf{q}}(k_c q^4 + \Gamma q^2) 
     z_{\mathbf{q}}^+ z_{\mathbf{-q}}^+\\
   & + 2(k_{\lambda} + \gamma_{\lambda}q^2) 
     \lambda_{\mathbf{q}}^+ \lambda_{\mathbf{-q}}^+\\
  +& \frac{1}{2} \sum_{\mathbf{q}}
     (k_A/t_0^2 + k_cq^4 - 4k_c\zeta q^2/t_0)
     z_{\mathbf{q}}^- z_{\mathbf{-q}}^-\\
   & + 2(k_{\lambda} + \gamma_{\lambda} q^2)
     \lambda_{\mathbf{q}}^- \lambda_{\mathbf{-q}}^- \quad .
\end{split}
\end{equation}
Here $z^{\pm}$ denotes the bending modes defined above, and $\lambda^{\pm} =
(\lambda_1 \pm \lambda_2)/2$ the corresponding protrusion modes. The parameters
$k_c$ and $k_A$ are the bending and compressibility moduli of the bilayer,
$\zeta$ is related to the spontaneous curvature $c_0$ and given by $c_0 -
\Sigma \: \textrm{d}c_0/\textrm{d} \Sigma$ with the area per lipid $\Sigma$,
$t_0$ is the mean monolayer thickness, and the parameters $\gamma_{\lambda}$ 
and $k_{\lambda}$ characterize the protrusion modes. 
The resulting spectra for height and thickness fluctuations are given by
\begin{equation}
\label{eq:elastic_height}
\langle | h_{\mathbf{q}} |^2 \rangle = 
\frac{k_B T}{k_c q^4 + \Gamma q^2} 
+ \frac{k_B T}{2(k_{\lambda} + \gamma_{\lambda}q^2)}\\
\end{equation}
\begin{equation}
\label{eq:elastic_width}
\langle | t_{\mathbf{q}} |^2 \rangle = 
\frac{k_B T}{k_c q^4 - 4 k_c \zeta q^2 / t_0 + k_A/t_0^2} 
+ \frac{k_B T}{2(k_{\lambda} + \gamma_{\lambda}q^2)} \quad .
\end{equation}

\begin{figure}
\begin{center}
\includegraphics[angle=0,width=\columnwidth]{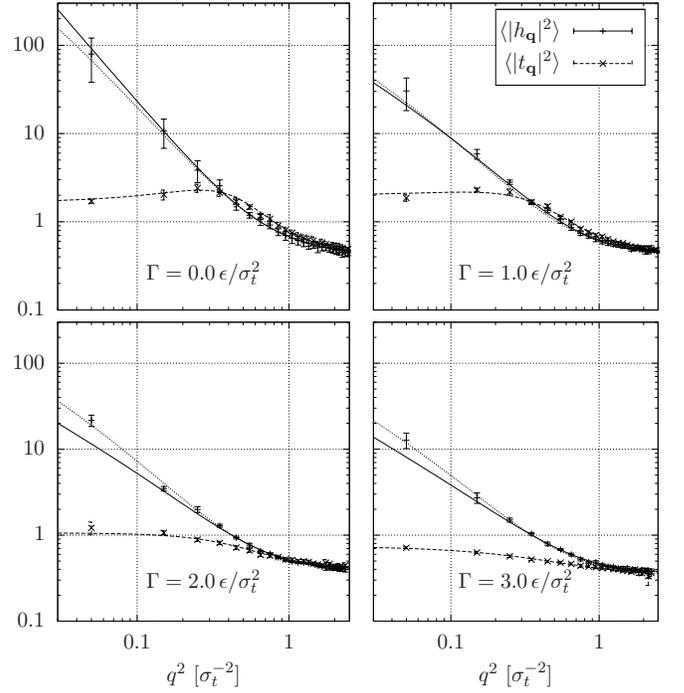}
\caption{
Fluctuation spectra of membranes for different tensions as
indicated. The dashed line is the fit to the thickness spectrum.  The
solid line shows the fit of the elastic theory to the height spectrum
for fixed tension $\Gamma$ and the dotted line shows the fit with
$\Gamma_{\textrm{fit}}$ as additional fit parameter.  At higher
tensions these fits deviate from each other. The values of
$\Gamma_{\textrm{fit}}$ are listed in
Table~\ref{tab:elastic_parameters_var_gamma}.
}
\label{fig:T1p3_40x40_g0-3-fluc_fit}
\end{center}
\end{figure}

We have determined the fluctuation spectra from simulations of systems
containing 3,200 lipids and 24,615 solvent beads using a method described
elsewhere \cite{loimar:03,wesbro:09}, and used the theory above to fit our
data. As in our earlier work \cite{wesbro:09}, the fitting yields very good
results for tensionless membranes. It still works well at the comparatively low
tension of $\Gamma = 1.0\,\epsilon/\sigma_t^2$.  In the high-tension regime
above  $2.0\,\epsilon/\sigma_t^2$ , however, the situation changes. When fixing
$\Gamma$ to the externally applied value, the fit consistently underestimates
the amplitudes of the long-wavelength fluctuations. This can be remedied by
leaving $\Gamma$ as fit parameter. However, the resulting values for the fitted
tension are smaller by a factor of up to one half than the externally applied
values (Fig.~\ref{fig:T1p3_40x40_g0-3-fluc_fit}).  The fit parameters for the
fits with fixed and free parameter $\Gamma$ are given in Tables
\ref{tab:elastic_parameters} and \ref{tab:elastic_parameters_var_gamma}.

\begin{table}
\begin{center}
\begin{tabular}{l c c}
$\Gamma$ [$\epsilon/\sigma_t^2$] 	& 0.0			& 1.0\\
\hline
\hline
$k_c$ [$\epsilon$]		 	& $4.9\pm1.0$		& $6.2 \pm 0.5$\\ 
\hline
$\zeta/t_0$ [$\sigma_t^{-2}$]	 	& $0.12\pm0.02$ 	&$0.069\pm0.019$\\
\hline
$k_A/t_0^2$ [$\epsilon/\sigma_t^4$] 	& $1.1\pm0.2$ 		& $0.92\pm0.09$\\ 
\hline
$k_{\lambda}$ [$\epsilon/\sigma_t^4$] 	& $1.4\pm0.1$ 		& $1.4\pm0.1$\\ 
\hline
$\gamma_{\lambda}$ [$\epsilon/\sigma_t^2$] 	& $0.022\pm0.039$ 		& $0.035\pm0.021$\\ 
\vspace{0.5cm}\\
$\Gamma$ [$\epsilon/\sigma_t^2$]	& 2.0 			& 3.0 \\
\hline
\hline
$k_c$ [$\epsilon$]			& $7.0 \pm 0.3$		& $7.7 \pm 0.3$\\
\hline
$\zeta/t_0$ [$\sigma_t^{-2}$]		& $-0.038\pm0.008$	&$-0.137\pm0.040$\\
\hline
$k_A/t_0^2$ [$\epsilon/\sigma_t^4$] 	& $1.8\pm0.2$		& $4.0\pm0.4$\\
\hline
$k_{\lambda}$ [$\epsilon/\sigma_t^4$] 	& $1.8\pm0.1$		& $2.0\pm0.1$\\
\hline
$\gamma_{\lambda}$ [$\epsilon/\sigma_t^2$]	& $-0.048\pm0.042$		& $-0.062\pm0.048$\\
\end{tabular}
\end{center}
\caption{Elastic parameters of the model membrane in the fluid phase,
obtained from a fit of the elastic theory (Eqs.~\ref{eq:elastic_height}
and \ref{eq:elastic_width}) with fixed $\Gamma$.
\\
}
\label{tab:elastic_parameters}
\end{table}

\begin{table}
\begin{center}
\begin{tabular}{l c c }
$\Gamma$ [$\epsilon/\sigma_t^2$] 	& 0.0			& 1.0\\		
\hline
\hline
$k_c$ [$\epsilon$]		 	& $4.4\pm0.7$		& $6.6 \pm 0.9$\\ 
\hline
$\zeta/t_0$ [$\sigma_t^{-2}$]	 	& $0.11\pm0.02$		&$0.074\pm0.019$\\
\hline
$k_A/t_0^2$ [$\epsilon/\sigma_t^4$] 	& $1.0\pm0.2$ 		& $0.95\pm0.11$\\
\hline
$k_{\lambda}$ [$\epsilon/\sigma_t^4$] 	& $1.5\pm0.1$ 		& $1.4\pm0.1$\\ 
\hline
$\gamma_{\lambda}$ [$\epsilon/\sigma_t^2$] 	& $0.018\pm0.043$ 		& $0.041\pm0.026$\\ 
\hline
$\Gamma_{\textrm{fit}}$ [$\epsilon/\sigma_t^2$] 	& $0.11\pm0.19$ 	& $0.83\pm0.14$\\ 

\vspace*{0.5cm}\\
$\Gamma$ [$\epsilon/\sigma_t^2$] 	& 2.0 			& 3.0 \\
\hline
\hline
$k_c$ [$\epsilon$]			& $10.9 \pm 0.8$        & $12.6 \pm 1.1$\\
\hline
$\zeta/t_0$ [$\sigma_t^{-2}$]		& $-0.011\pm0.012$	&$-0.11\pm0.08$\\
\hline
$k_A/t_0^2$ [$\epsilon/\sigma_t^4$]	& $2.0\pm0.3$		& $4.3\pm0.5$\\
\hline
$k_{\lambda}$ [$\epsilon/\sigma_t^4$]	& $1.6\pm0.03$		& $1.7\pm0.1$\\
\hline
$\gamma_{\lambda}$ [$\epsilon/\sigma_t^2$]& $0.057\pm0.030$ 	        & $0.048\pm0.032$\\
\hline
$\Gamma_{\textrm{fit}}$ [$\epsilon/\sigma_t^2$] & $0.90\pm0.31$ 	& $1.7\pm0.4$\\
\end{tabular}
\end{center}
\caption{Elastic parameters of the model membrane in the fluid phase from
a fit to Eqs.~\ref{eq:elastic_height} and \ref{eq:elastic_width} where
the parameter $\Gamma$ is allowed to vary. The fitted value is
$\Gamma_{\textrm{fit}}$.
}
\label{tab:elastic_parameters_var_gamma}
\end{table}

The relation between external ($\Gamma$) and internal stress
($\Gamma_{\textnormal{int}}$) in membranes has been discussed by a number of
authors for the situation where the membrane is kept in a frame with fixed
projected area $A$. In these studies, the {\em total area} $A_s$ was allowed to
fluctuate, either because of fluctuations in the number of molecules
(grandcanonical case) \cite{cailub:94, farpin:03} or because of fluctuations of
the lipid area (canonical, compressible case) \cite{farpin:03, imp:06}. The
frame tension is then found to differ from the internal stress in the membranes
due to the contribution of the membrane fluctuations to the surface free energy
\cite{brogen:76,davlei:91,cailub:94,farpin:03,imp:06}.  The correction is
additive and should always be present, even at (external or internal) tension
zero. For the canonical, compressible case, which is obviously more relevant in
our context, Farago \emph{et al.} \cite{farpin:03} and Imparato \cite{imp:06}
have predicted that the fluctuations reduce the frame tension by roughly
$k_BT\,n/A$, compared to the intrinsic stress, where $n$ is the number of
fluctuation degrees of freedom, {\em i.e.}, the number of independently
fluctuating membrane patches. Thus the intrinsic tension should be higher than
the frame tension, which is opposite from what we observe in our simulations.

However, we believe that the two situations -- fixed frame and varying surface
area versus variable frame and (roughly) fixed surface area -- are not
comparable.  According to the arguments leading to Eqs.~(\ref{eq:gamma_1}) and
(\ref{eq:gamma_2}), the frame tension is {\em not} renormalized by fluctuations
at the level of a Gaussian theory ({\em i.e.}, a theory based on a free energy
functional which is quadratic in the fluctuations). It might be renormalized if
one includes higher order terms. For example, the last term in
Eq.~(\ref{eq:gamma_2}), $ \propto \int \mathrm{d}^2 r \: \delta, $ introduces a
thickness-mediated interaction between the height fluctuation modes {\em via}
the relation
\begin{equation}
\delta = 2 z_- / \sqrt{1 + (\nabla z_+)^2} - \overline{D}
\approx 2z_- - z_- \: (\nabla z_+)^2 \, - \overline{D},
\end{equation}
which might effectively renormalize $\Gamma_{\textnormal{int}}$.
Another possibility is of course that the theory of Brannigan and
Brown \cite{brabro:06}, which we have used to analyze the data, is no
longer applicable at high tensions due to the structural
rearrangements in the membrane.

\subsection{Lipid-mediated interactions between inclusions}
\label{sec:pmf}

Finally in this section, we discuss the effect of tension on the
membrane-mediated interactions between two simple cylindrical inclusions in the
bilayer. We focus on the effective interactions between these model proteins
and the influence of an external tension on the potential of mean force (PMF).
The radial distribution function $g(r)$ as a function of the protein-protein
distance $r$ was obtained from simulation runs using the technique of
successive umbrella sampling \cite{virmue:04} combined with a reweighting
procedure. As starting configurations we used equilibrated systems with 750 to
760 lipids and two simple transmembrane proteins of diameter $3\,\sigma_t$. A
first estimate of $g(r)$  was obtained during $2\times10^6\,\MCS$.  Then,
biased runs of $3\times10^6\,\MCS$ were performed to improve the statistics of
configurationally less frequent protein-protein distances. After removing the
bias from these results and combining the overlapping distributions the
effective potential $w(r) = -k_B T \ln g(r)$ was extracted. In order not to
complicate the interpretation of our results, the model proteins were not
allowed to tilt. This can be justified by assuming that real transmembrane
proteins might be bound to, e.g.  cytoskeletal, structures outside the
membrane, which allow for transverse motion but not for tilt.

The type of model for the inclusions is identical to the one introduced in
\cite{wesbro:09} and a brief overview is given in the following: The
interaction of this simple model protein and the lipid or solvent beads has a
repulsive contributions, which is described by a radially shifted and truncated
Lennard-Jones potential
\begin{equation}
V_{\textrm{rep}} (r_{xy}) = \left\{
\begin{array}{rl}
V_{\textrm{LJ}} \left( \frac{r_{xy} - \sigma_0}{\sigma} \right) -
V_{\textrm{LJ}}(1) & 
\textrm{if $r_{xy} - \sigma_0 < 0$}\\
0 & \textrm{otherwise}\\
\end{array}  \right .
\label{equ:V_rep}
\end{equation}
where $r_{xy} = \sqrt{x^2 + y^2}$ denotes the distance of the interaction
partners in the $xy$ plane, $\sigma$ is given by $\sigma = (\sigma_t
+ \sigma_i)/2$ for interactions with beads of type $i$ ( $i = h$, $t$,
and $s$ for head, tail, and solvent beads, respectively), $\sigma_0 =
\sigma_t$, and $V_{\textrm{LJ}}$ has been defined above
(Eq.~\ref{equ:V_LJ}). The direct protein-protein interactions have the
same potential (Eq.~\ref{equ:V_rep}) with $\sigma = \sigma_t$ and
$\sigma_0 = 2\sigma_t$.

In addition, protein cylinders attract tail beads on a hydrophobic
section of length $L$. This is described by an additional attractive
potential that depends on the $z$ distance between the tail bead and
the protein center. The total potential reads
\begin{equation}
V_{pt}(r_{xy}, z) = \epsilon_{pt} \left(V_{\textrm{rep}}(r_{xy}) +
V_{\textrm{attr}}(r_{xy}) \times W_p(z) \right)\, ,
\label{equ:V_pt}
\end{equation}
with the attractive Lennard-Jones contribution
\begin{equation}
V_{\textrm{attr}} = \left\{
\begin{array}{rl}
V_{\textrm{LJ}}(1) - V_{\textrm{LJ}}(2) & 
\textrm{if $r_{xy} - \sigma_0 < \sigma$}\\
V_{\textrm{LJ}}\left( \frac{ r_{xy} - \sigma_0}{\sigma} \right) - V_{\textrm{LJ}}(2) &
\textrm{if $\sigma < r_{xy} - \sigma_0 < 2 \sigma$}\\
0 & \textrm{otherwise}
\end{array} \right . ,
\label{equ:V_attr}
\end{equation}
and a weight function $W_p(z)$, which is unity on a stretch of length
$2 l =  L - 2 \sigma_t$ and crosses smoothly over to zero over a
distance of approximately $\sigma_t$ at both sides. Specifically, we
use
\begin{equation}
W_p = \left \{
\begin{array}{rl}
1 & \textrm{if $|z| \leq l$}\\
\cos^2 \left( \frac{3}{2}|z| - l \right) & \textrm{if $l < |z| < l +
\frac{\pi}{3}$}\\
0 & \textrm{otherwise}
\end{array} \right. .
\label{equ:W_P}
\end{equation}
The hydrophobicity of the protein is tuned by the parameter
$\epsilon_{pt}$. The choice of a sufficiently high
interaction  strength between the hydrophobic core of the membrane and
the hydrophobic part of the inclusion is crucial to induce local
perturbation of the bilayer. We note that the repulsive sections of
our model proteins span the whole simulation box in $z$ direction.
Therefore, the simulations were carried out at constant box height 
$L_z$, volume moves were only allowed in the lateral directions 
$x$ and $y$, and the number of solvent beads of the was allowed to 
fluctuate (see Sec.~\ref{sec:simulation}).

\begin{figure}
\begin{center}
\includegraphics[angle=0,width=\columnwidth]{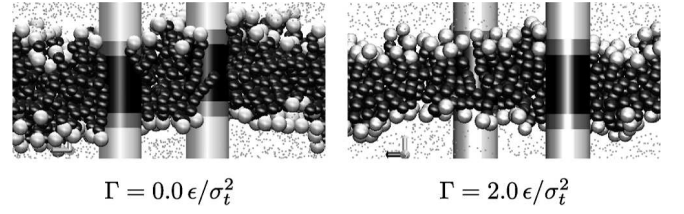}
\caption{Vertical slices through systems with two inclusions. The dark gray
rings above the black hydrophobic part of the protein indicate the sections
where the attractive interaction between protein and tail beads (dark gray)
decays to zero. The light gray section and the light gray beads mark the
hydrophilic part of the protein and the lipid heads, respectively.  The solvent
beads above and below the bilayer are marked by small dots.  The left snapshot
shows a stressfree bilayer with proteins of hydrophobically matching length $L
= 6\,\sigma_t$. On the right the same proteins are shown in a bilayer under
a tension of $\Gamma = 2.0\,\epsilon/\sigma_t^2$.}
\label{fig:T1p3_umbrella_snapshots}
\end{center}
\end{figure}

We found that the effect of tension on the PMF was only significant for rather
hydrophobic model proteins, {\em i.e.}, proteins with high interaction
parameter $\epsilon_{pt}$.  In the following, we will present the
results obtained with $\epsilon_{pt} = 6.0\,\epsilon$.
Fig.~\ref{fig:T1p3_umbrella_snapshots} shows snapshots of the model proteins 
in membranes at zero tension and at tension 
$\Gamma = 2.0 \, \epsilon/\sigma_t^2$.

\begin{figure}
\begin{center}
\includegraphics[angle=0,width=\columnwidth]{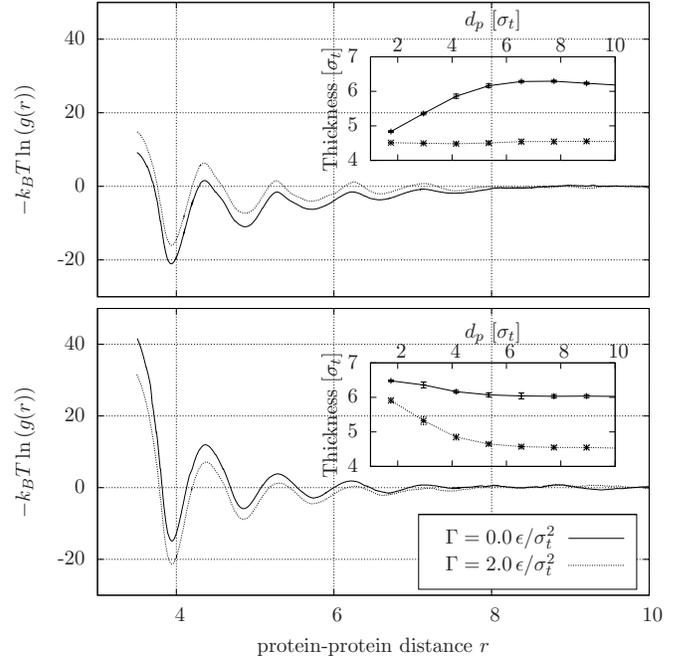}
\caption{
Potential of mean force of two isolated proteins as a function of the
protein distance at tension zero (solid lines) and 
$\Gamma = 2.0\,\epsilon/\sigma_t^2$ (dotted lines), for proteins
with hydrophobic length $L = 4\,\sigma_t$ (top) and 
$L = 6\,\sigma_t$ (bottom).  The insets show the corresponding radial 
thickness profiles around a single protein ($d_p$ denotes the in-plane 
distance to the center of the inclusion).
}
\label{fig:T1p3_pl2-4_umbrella}
\end{center}
\end{figure}

The potentials of mean force between two inclusions as a function of their
in-plane distance are plotted in Fig.~\ref{fig:T1p3_pl2-4_umbrella} for two
different hydrophobic lengths $L$. In the first case
(Fig.~\ref{fig:T1p3_pl2-4_umbrella}, top), the inclusion is characterized by a
negative hydrophobic mismatch in tensionless free membranes. Under tension, the
membrane thins and the mismatch is reduced (inset
Fig.~\ref{fig:T1p3_pl2-4_umbrella}, top).  In the second case
(Fig.~\ref{fig:T1p3_pl2-4_umbrella}, bottom), the hydrophobic part of the
inclusion roughly matches the thickness in the tensionless case. Under tension,
the membrane thins and a positive hydrophobic mismatch develops (see inset
Fig.~\ref{fig:T1p3_pl2-4_umbrella}, bottom). 

Due to lipid packing in the vicinity of the inclusions the curves show an
oscillatory shape with a wavelength of approximately $1\,\sigma_t$.  Since at
small inclusion-inclusion distances direct interactions of the proteins  and
depletion induced attraction due to the solvent particles come into play, these
parts of the curves have been cut off. Thus, we can focus on the lipid-mediated
medium and long ranged interactions. The main effect of tension on the lipid
mediated interactions between the two model proteins can be summarized as
follows: In the absence of tension, the interactions between hydrophobically
mismatched inclusions have an additional attractive contribution, compared to
hydrophobically matched inclusions. In the case where the tension reduces the
hydrophobic mismatch due to membrane thinning, this attraction diminishes. If,
on the other hand, the tension leads to a stronger hydrophobic mismatch, the
attractivity of the interaction potential also increases. Therefore, we
conclude that the dominant effect of an external tension on the lipid-mediated
interactions is indirect and related to the change of the hydrophobic mismatch
due to the membrane thinning. For decreasing (negative) mismatch, the average
attraction decreases, and for increasing (positive) mismatch, it increases. This
is consistent with the behavior observed in tensionless membranes, where also
both positive and negative hydrophobic mismatch resulted in an attractive
contribution to the PMF \cite{wesbro:09}.  Other recent studies
\cite{demven:08,schgui:08,demsmi:09:2,schgui:09} have also highlighted the
importance of hydrophobic mismatch as a driving factor for protein
agglomeration. 
From our simulations, an additional effect of tension is not evident.


\section{Discussion and Summary}

In this paper we have studied the influence of an external tension on the
properties of bilayers using a generic coarse-grained model. To put these
results into perspective, we will now briefly discuss the experimental 
situation.

Experimentally, one of the most widely used techniques to determine mechanical
stretch properties of bilayers is the micropipette approach, where giant
bilayer vesicles are pressurized by micropipette suction \cite{kwoeva:81}. This
method produces a uniform membrane tension, is very accurate, and can be used
to verify elastic reversibility \cite{rawolb:00}.  Micropipette aspiration
experiments, {\em e.g.}, carried out by Needham and Nun \cite{neenun:90}, found
for different lipids and lipid/cholesterol mixtures that membrane lysis is
usually reached at an relative areal expansion of less than $5\,\%$, and the
rupture strength at low cholesterol concentration was typically around
$2 - 10\,\textnormal{mN/m}$.

This does not compare well with our findings and those of other atomistic or
mesoscopic simulations (see the introduction to Sec.~\ref{sec:results}), where
fluid membranes could sustain tensions of $30\,\textnormal{mN/m}$ or more and
remained stable up to relative extensions of 40 \% or more.  It should however
be noted that tension-induced lysis is a stochastic process, and the duration
of the exposure to the stress plays a decisive role for the stability.
Experimentally, abrupt failure of the bilayer is observed, when the tension
reaches a critical value, whereas below this tension long-term persistence of
the stressed membrane can be presumed \cite{evanee:87}. Evans \emph{et al.}
showed that the rupture strength of membranes is a property which crucially
depends on the loading rate \cite{evahei:03}. In their study on five types of
fluid giant phosphatidylcholine lipid vesicles, they varied the loading rate
from $0.01 - 100\,\textnormal{mN/m/s}$. At high loading rates the systems were
found to be stable up to values of $20 - 30\,\textnormal{mN/m}$.

To get a rough estimate of the meaning of the timescales in our simulations,
compared to experimental systems, we can map the diffusion constants $D_{xy}$
in Sec.~\ref{sec:diffusivity} to the corresponding values measured for DPPC
bilayers in recent experiments.  At $45^{\circ}\,\textrm{C}$ Scomparin \emph{et
al.} \cite{scolec:09} report a diffusion coefficient of approximately
$10\,\mu\textrm{m}^2/\textrm{s}$.  Taking our result of $D_{xy} = 0.8 \times
10^{-5}\,\sigma_t^2 /\textrm{MCS}$ for the tensionless membrane and setting our
intrinsic length scale $1\,\sigma_t$ to $6\,\textrm{\AA}$ as described earlier,
we find that $10^6\,\textrm{MCS}$ in our simulation correspond to approximately
$0.3\,\mu\textrm{s}$ in real time. The typical lengths of our simulations
lie between $2$--$4 \times 10^6\,\textrm{MCS}$, which corresponds to roughly
$1\,\mu\textrm{s}$ in real systems. Thus our simulations correspond to systems
exposed to very high loading rates and short timescales. Taking this into
consideration, their stability does not contradict experimental findings.
Our simulations provide a way to study membranes under extreme conditions, 
and to analyze their structural properties, which cannot be accessed easily
by experiments.

Due to these difficulties, experimental results with which we could compare our
simulations are scarce. One positive example is the behavior of the the ripple
phase under stress. To our knowledge, we have performed the first simulation
study which tries to shed light on the structural rearrangements and elastic
properties of a bilayer in the ripple phase under lateral stress.  We have
shown that lateral tension leads to suppression of the ripple structure in the
$P_{\beta'}$ phase, and we see a transition of the areal extensibility from
soft, fluid-like to gel-like behavior.  Qualitatively, this behavior agrees
well with the findings of Evans \emph{et al.} \cite{evanee:87, neemci:88}, who
also reported an initial soft-elastic response at low tensions and stiff
elastic properties after elimination of the ripple.  

In the gel phase, the tension does not change the state of the bilayer
significantly in the range of tensions considered in this work. The basic
structure of the two monolayers stays intact.  The situation is very different
in the biologically most relevant fluid phase. At temperatures where the
tensionless membranes are fluid, they respond to high tensions by a structural
change from a state where both monolayers are well-separated to a state where
they are partly interdigitated. These changes are associated with substantial
variations of the compressibility  (up to a factor of 3), and the lipid
diffusion constant (up to 20 \%). 

We have also studied the influence of membrane tension on the effective
interaction between two model proteins. Under tension, the membrane becomes
thinner, which affects the hydrophobic mismatch interaction. This is found to
be the dominant effect. The interaction between negatively mismatched proteins
decreases, and that between positively mismatched proteins increases. Thus
applying tension can be used to tune the strength of membrane 
mediated protein-protein interactions.

\section*{Acknowledgement}

Provision of computing resources by the HLRS (Stuttgart), NIC
(J\"ulich), and PC2 (Paderborn) is gratefully acknowledged.
The configurational snapshots were visualized using VMD \cite{humdal:96}. 
This work was funded by the DFG within the Sonderforschungsbereich
SFB 613 and SFB 625.


\end{document}